\documentclass[notitlepage,12pt]{article}
\usepackage{latexsym}
\usepackage[T1]{fontenc}

\begin{document}
\textwidth=5in \textheight=7.5in \pagestyle{myheadings}
\makeatletter
\pagenumbering{arabic}


\title{Relativistic Quantum Mechanics and \\ Field Theory \thanks
{Invited lecture at the Conference "Problemi Attuali di Fisica
Teorica", Sessione: Meccanica Quantistica ("Present Problems of
Theoretical Physics", Session: Quantum Mechanics), IIASS "E.R.
Caianiello", Vietri sul mare (SA), Italy, 11-16 April, 2003.}}
\author{F. Strocchi
\\  Scuola Normale Superiore and INFN, Pisa, {\em Italy}}

\date{}

\maketitle

\makeatletter \@addtoreset{equation}{section}

\vspace{10mm}
The problems which arise for a relativistic
quantum mechanics are reviewed and critically examined in
connection with the foundations of quantum field theory. The
conflict between the quantum mechanical Hilbert space structure,
the locality property  and the gauge invariance encoded in the
Gauss' law is discussed in connection with the various
quantization choices for gauge fields.

\vspace{3mm} {\bf KEY WORDS:} relativistic quantum mechanics,
quantization of gauge theories


\newtheorem{Theorem}{Theorem}[section]
\newtheorem{Definition}{Definition}[section]
\newtheorem{Proposition}{Proposition}[section]

\def\theequation{\thesection.\arabic{equation}}

\def \eqq {\equiv}
\def \Pf {{\bf Proof}\,\,\,\,}
\def \Rbf {{\bf R}}
\def\Ra{\Rightarrow}
\def\ra{\rightarrow}
\def\Id{\mathop{\bf Id}}
\def\dalamb{{\sqro86}{\hskip1.1pt}}
\@addtoreset{equation}{section}
\def\AO'{\mbox{${\cal A}({\cal O}')$}}
\def\O{\mbox{${\cal O}$}}
\def \cal {\mathcal}
\def \o {{\omega}}
\def \O {{\mathcal O}}
\def \OM {{\Omega}}
\def \A {{\mathcal A}}
\def \AO {\A(\O)}
\def \B {{\cal B}}
\def \F {{\cal F}}
\def \D {{\cal D}}
\def \H {{\cal H}}
\def \K {{\cal K}}
\def \P {{\cal P}}
\def \L {{\cal L}}
\def \S {{\cal S}}
\def \Sp {{{\cal S}^\prime}}
\def \Z  {{\cal Z}}
\def \W  {{\cal W}}

\def \psz {{\Psi_0}}
\def \psb {{\bar\psi}}
\def \a {\alpha}
\def \b {\beta}
\def \d {\delta}
\def \l {\lambda}
\def \t {\tau}
\def \g {{\gamma}}
\def \Cf {{C^\infty}}

\def\naturali{{\bf N}}
\def \rep {representation }
\def \reps {representations }
\def \ip {inner product }
\def \nd {non--degenerate }
\def \qft {quantum field theory }
\def \ex {extension }
\def \exs {extensions }
\def \wcs {weakly compatible sequences }
\def \ucs {u-weakly compatible sequences }

\def \eps {{\varepsilon}}
\def \Dz {{{\cal D}_0}}
\def \Da {{{\cal D}_\alpha}}
\def \Iz {{{\cal I}_0}}
\def \Iza {{{\cal I}_0^\alpha}}
\def \deg {<x,y> \ = 0, \ \ \forall y \in \Dz}
\def \fayz {\forall y \in \Dz}
\def \tw {{\tau_w}}
\def \pyx {{{p_y (x)}}}
\def \ps {{ < {\cdot} , {\cdot} > }}
\def \psnm {{ < x_n , y_m > }}
\def \xn { \{ x_n \} }
\def \yn { \{ y_n \} }
\def \xnx {{x_n \to x}}
\def \yny {{y_n \to y}}
\def \lnm {{\lim_n \lim_m}}
\def \lmn {{\lim_m \lim_n}}
\def \LA  {\Lambda}
\def \Ta {{T_\alpha }}
\def \Sa {{S_\alpha }}
\def \ux {{\underline x}}
\def \uy {{\underline y}}
\def \uz {{\underline z}}
\def \uf {{\underline f}}
\def \xnDz {{x_n \in \Dz}}
\def \dm {{\partial_\mu}}
\def \dn {{\partial_\nu}}

\def \di {{\partial_i}}
\def \dj {{\partial_j}}
\def \dk {{\partial_k}}
\def \dt {{\partial_t}}
\def \de {{\partial}}

\def \be {\begin{equation}}
\def \e {\end}
\def \yyyyy {\end{equation}}
\def \x {{\bf x}}
\def \y {{\bf y}}
\def \z {{\bf z}}
\def \k {{\bf k}}
\def \SR {\S(\Rbf^3)}
\def \DR {\D(\Rbf^3)}
\def \psio {\Psi_0}

\def \fb {\overline{f}}
\def \gb {\overline{g}}
\def \df {\partial f}
\def \dg {\partial g}

\def \be {\begin{equation}}
\def \ee {\end{equation}}
\def \ume {{\scriptstyle{\frac{1}{2}}}}
\def \ra {\rightarrow}
\def \Ra {\Rightarrow}
\def \eqq {\equiv}

\def \a {{\alpha}}
\def \b {{\beta}}
\def \g {{\gamma}}
\def \d {{\delta}}
\def \eps {{\varepsilon}}
\def \th {{\theta}}
\def \l {{\lambda}}
\def \La {{\Lambda}}
\def \s {{\sigma}}
\def \Si {{\Sigma}}
\def \t {{\tau}}
\def \ph {{\varphi}}
\def \phb {{\overline{\varphi}}}
\def \o {{\omega}}
\def \Om {\mbox{${\Omega}$}}
\def \Ga {{\Gamma}}

\def \A {{\cal A}}
\def \B {{\cal B}}
\def \C {{\cal C}}
\def \D {{\cal D}}
\def \F {{\cal F}}
\def \G {{\cal G}}
\def \H {\mbox{${\cal H}$}}
\def \J {{\cal J}}
\def \K {{\cal K}}
\def \L {{\cal L}}
\def \N {{\cal N}}
\def \O {{\cal O}}
\def \P {{\cal P}}
\def \S {{\cal S}}
\def \U {{\cal U}}
\def \V {{\cal V}}
\def \W {{\cal W}}
\def \Z {{\cal Z}}

\def \Id {\mathop{\bf Id}}
\def \id {{\bf 1 }}
\def \eijk {{\varepsilon_{ijk}}}
\def \eklm {{\varepsilon_{klm}}}
\def \dij {{\delta_{ij}}}
\def \dik {{\delta_{ik}}}
\def \djk {{\delta_{jk}}}
\def \dkl {{\delta_{kl}}}
\def \Psio {{\Psi_0}}

\def \di {{\partial_i}}
\def \dj {{\partial_j}}
\def \dk {{\partial_k}}
\def \dl {{\partial_l}}
\def \do {{\partial_0}}
\def \dz {{\partial_z}}

\def \dmu {{\partial_\mu}}
\def \dnu {{\partial_\nu}}
\def \dla {{\partial_\lambda}}
\def \dr {{\partial_\rho}}
\def \ds {{\partial_\sigma}}
\def \dt {{\partial_t}}
\def \do {{\partial_0}}
\def \dum {{\partial^\mu}}
\def \dun {{\partial^\nu}}
\def \Amu {{A_\mu}}
\def \Anu {{A_\nu}}
\def \Fmn {{F_{\mu\,\nu}}}
\def \jm  {j_\mu}

\def \abf {{\bf a}}
\def \bbf {{\bf b}}
\def \cbf {{\bf c}}
\def \hbf {{\bf h}}
\def \k {{\bf k}}
\def \kbf {{\bf k}}
\def \jbf {{\bf j}}
\def \j   {{\bf j}}
\def \nbf {{\bf n}}
\def \q {{\bf q}}
\def \qbf {{\bf q}}
\def \p {{\bf p}}
\def \pbf {{\bf p}}
\def \sbf {{\bf s}}
\def \rbf {{\bf r}}
\def \ubf {{\bf u}}
\def \vbf {{\bf v}}
\def \xbf {{\bf x}}
\def \x {{\bf x}}
\def \y {{\bf y}}
\def \ybf {{\bf y}}
\def \v {{\bf v}}
\def \z {{\bf z}}
\def \zbf {{\bf z}}
\def \Rbf {{\bf R}}
\def \Cbf {{\bf C}}
\def \Nbf {{\bf N}}
\def \Zbf {{\bf Z}}
\def \Abf {{\bf A}}
\def \Jbf {{\bf J}}

\newcommand{\mbf}[1] {\mbox{\boldmath{$#1$}}}

\def \AO {{\cal A}({\cal O})}
\def \AO' {{\cal A}({\cal O}')}
\def \Aob {\A_{obs}}
\def \dxy {\delta(x-y)}
\def \at {{\alpha_t}}
\def \ax {{\alpha_{\x}}}
\def \atv {{\alpha_t^V}}

\def \fR {{f_R}}
\def \hf {\tilde{f}}
\def \tilf {\tilde{f}}
\def \tilg {\tilde{g}}
\def \tilh {\tilde{h}}
\def \tilF {\tilde{F}}
\def \tilJ {\tilde{J}}

\def \cc  {\subseteq}
\def \] {\supseteq}

\def \pio {{\pi_\o}}
\def \pom {{\pi_{\Omega}}}
\def \Hom { {\H_{\Omega}} }
\def \Psiom  { \Psi_{\Omega} }

\def \Pf {{\bf Proof.\,\,}}
\def \limx {{\lim_{|\x| \ra \infty}}}
\def \frx {f_R(x)}
\def \limR {\lim_{R \ra \infty}}
\def \limV {\lim_{V \ra \infty}}
\def \limko {\lim_{k \ra 0}}

\def \Roo {R \ra \infty}
\def \ko {k \ra 0}
\def \jo {j_0}
\def \su {{ \left(
\begin{array}{clcr} 0 & 1 \\1 & 0 \end{array} \right)}}
\def \sd {{ \left(
\begin{array}{clcr} 0 & -i \\i & 0 \end{array} \right)}}
\def \st {{ \left(
\begin{array}{clcr} 1 & 0 \\0 & -1 \end{array} \right)}}

\maketitle


\baselineskip=6mm
\section{INTRODUCTION}
The problems of relativistic quantum mechanics (RQM) may look (at
least partly)  obsolete, but since the  solution offered by
quantum field theory (QFT) is now challenged by more radical ideas
(like string theory and/or non commutative space time), it is
perhaps worthwhile to critically reexamine the conceptual
motivations for the birth of QFT. The difficulties of RQM which
QFT is supposed to bypass are mentioned in almost any book on QFT,
but in our opinion, in view of the latest developments, it may be
useful to focus the basic properties which enter in the game and
the role played by covariance, spectral condition (or stability),
relativistic covariance and locality in forcing the transition
from one particle (or few particles) Schroedinger QM to quantum
fields.

Another conceptual reason for  a reexamination of the foundational
problems at the origin of QFT is that after more than seven
decades  no non trivial (even non realistic) model in four (space
time) dimensions is under non perturbative control. Actually, the
prototypic model of self interacting scalar field, which is used
in most textbooks for developing (non trivial) perturbation
theory, has been proved to be trivial (namely the renormalized
coupling constant vanishes when the ultraviolet cutoff is removed)
under general conditions, when treated non perturbatively
\footnote{For a review of the arguments on the triviality of
$\ph^4$ theories see Ref. 1.}. This means that in general the
perturbative expansion is not reliable and in general one cannot
define a QFT model by its perturbative expansion.

It should be stressed  that most of our wisdom in QFT is derived
from the perturbative expansion and that it would be silly to
neglect the extraordinary success of perturbative Quantum
Electrodynamics (QED) in yielding  theoretical predictions which
agree with the experiments up to the eleventh significant figure.
On the other hand, soon after the setting of perturbation theory,
Dyson argued $^{(2)}$ that the perturbative expansion of QED
cannot be summed and that big oscillations overwhelming the so
successful lowest orders are expected to arise (typically at order
$n = 1/\a = 137$).

These negative results legitimate the need of a non perturbative
approach to the problem of combining quantum mechanics and
relativity, with the aim of either validating the foundations of
quantum field theory or displaying the need of radical changes and
new ideas.\footnote{In the history of theoretical physics, there
are famous examples  in which the combination of  different
theories, with different origins, like e.g. electromagnetism and
thermodynamics or electromagnetism and mechanics, turned out to be
impossible without dramatic conceptual revolutions, like Planck
energy quanta and special relativity.}

It is a common belief that (non abelian) gauge theories provide
the way out of the triviality theorems, but again a non
perturbative control is lacking; moreover as we shall discuss
below, such theories involve strongly delocalized (field)
variables (typically those carrying a non zero charge), whose
quantization requires either non regular representations of the
canonical commutation relation relation (CCR) or a violation of
positivity by their vacuum correlation functions. In both cases,
the quantum mechanical interpretation of such variables is not
standard.

\sloppy
\section{QUANTUM MECHANICS AND RELATI\-VI\-TY}
\fussy
Soon after the birth of quantum mechanics it became clear
that in order to describe microscopic systems, like electrons,
protons etc., at high energies one should combine quantum
mechanics and relativity. \def \do {\partial_0}

Both theories are very well sound and fully under control, in
their domain of applications, also from a mathematical point of
view; however, as we shall see, the combination of the two is a
non trivial problem.

The basic requirements one has to fulfill are the following
\newline ({\bf QUANTUM MECHANICS})  {\bf A Hilbert space
structure} for the description of the (physical) states of the
system, a {\bf unitary dynamics} $U(t)$ and a {\bf stability}
condition, i.e. a Hamiltonian bounded below. \newline({\bf
RELATIVITY})\, {\bf Relativistic covariance},  {\bf relativistic
spectral condition} $P^2 \geq 0, \,\,P_0  \geq 0$, and {\bf
locality} (or hyperbolic dynamics).

The implementability of  time evolution by (a strongly continuous
one parameter group of) unitary operators $U(t)$ is equivalent to
the existence of the Hamiltonian (as a self adjoint operator); the
relativistic spectral condition is actually required by the
relativistic invariance of the property of energy positivity.

An hyperbolic dynamics is necessary for a finite propagation
speed, as required by Einstein causality. Observable quantities
should in fact be localizable (in bounded regions) and their time
evolution should be contained in the corresponding causal shadow.
\sloppy
\subsection{Relativistic Schroedinger Quantum  Mechanics}
\fussy Historically, the first attempts to combine quantum
mechanics (QM) and relativity (R) went in the direction of writing
a relativistic version of the Schroedinger equation ({\em
relativistic wave equations}), and it was soon realized that
serious problems emerge. In particular, it turned out to be
impossible to satisfy basic physical properties, namely positivity
of the energy, locality  and non trivial interaction.

The simplest step is to replace the non relativistic
energy-momentum dispersion law $E = \p^2/2m$ by the relativistic
one $E = \sqrt{\p^2 + m^2}$ and obtain the relativistic
Schroedinger equation (for simplicity we chose units such that $
\hbar = c = 1 $)  \be{ i \dt \,\psi = \sqrt{ - \Delta + m^2 }\,
\psi. }\ee On the right hand side we have a pseudodifferential
operator, whose mathematical meaning  is that of acting as the
multiplication operator $\sqrt{\p^2 + m^2}$ on the Fourier
transform of $\psi$. Such an operator is non local and in fact it
implies that even if the initial data is of compact support, it
cannot remain so at later times.
\begin{Proposition} Equation (2.1) does not have solutions of
compact support.
\end{Proposition}
\Pf  In fact, if $\psi(\x, t)$ is of compact support for $t \in
[0, \eps)$ so  is its time derivative and the Fourier transforms
of both, with respect to $\x$, are analytic functions of $\p$ .
Now, by eq.(2.1), $\dt \tilde{\psi}(\p,t ) = - i \sqrt{\p^2 +
m^2}\,\tilde{\psi}(\p, t)$ and the discontinuity  of the square
root across the cut running from $- m^2$ to $\infty$ cannot be
removed by multiplication by the analytic function
$\tilde{\psi}(\p, t)$.

\vspace{2mm}The lack of locality of the time evolution is a
serious drawback. First, localizability of the wave functions is
strictly related to the possibility of implementing  relativistic
causality, namely the property that observable densities can be
localized so that their support evolves in time with velocity less
that $c = 1$. In particular, for solutions of eq.(2.1) the density
$\psi(\x)^* \psi(\x)$ cannot be localized in this sense. The same
localization problem arises for the current density $ j_\mu(\x) =
(j_0(\x), \,j_i(\x))$, \begin{eqnarray} j_0(\x) &=&
\ume\,[\psi^*(\x) \,(\sqrt{ - \Delta + m^2}\psi)(\x) + (\sqrt{ -
\Delta + m^2}\psi^*)(\x)\,\psi(\x)], \nonumber \\ j_i(\x) &=& \ume
i \,[ \psi^*(\x)\,\nabla_i\psi(\x) -
\nabla_i\psi^*(\x)\,\psi(\x)].\end{eqnarray} If $j_0$ is of
compact support at initial time, it does not remain so at any
later time. \footnote{Moreover, if the form of $j_0$ is used to
define the scalar product between two wave functions, the
hermiticity of operators is not the same as in Schroedinger QM; in
particular the multiplication by $\x$ is not hermitean and cannot
describe the position operator. An hermitean position operator
$$\x^{op} \eqq \x - \ume\, {\bf{\nabla}} (-\Delta + m^2)^{-1}$$
can be introduced, but the maximum localization has a tail of
exponential decay and such "localization" property is not stable
under Lorentz transformations.}

Furthermore, the non locality of the free Hamiltonian makes very
difficult to introduce the interaction, like the minimal coupling
with an electromagnetic potential, since it is hard to give a
simple meaning to the formal operator $\sqrt{(\p - e {\bf
A(\x)})^2 + m^2}$ (apart from the definition which requires the
knowledge of the spectrum of the operator under square root).

A local time evolution is given by the Klein-Gordon equation
\be{(\Box + m^2) \ph(x) = 0,}\ee which  is a hyperbolic equation
and therefore preserves the localization of the initial data with
a finite propagation speed. However, eq.(2.3) has solutions whose
Fourier transform with respect to time has support unbounded
below, i.e. the frequency $\o$  may take arbitrarily large values
with both signs $\o = \pm \sqrt{\k^2 + m^2}$; the negative sign
violate the positive energy spectral condition. For solutions with
negative frequencies also $\rho(\x, t) \eqq j_0(\x, t)$ becomes
negative and there is no good candidate for a probability density.
The solutions with negative frequencies must therefore be excluded
by means of a supplementary condition. Since eq.(2.3) is of second
order in the time derivative, the initial data involve both the
value of $\ph$ and of its time derivative and the energy spectral
condition requires that the initial data must satisfy the
condition
\def \do {\partial_0}
\be{i\, \do \ph(\x, 0) =  \sqrt{ - \Delta + m^2} \ph(\x, 0).}\ee
However, by the previous argument this is a non local condition
and therefore the initial data for solutions satisfying the
positive energy spectral condition cannot have compact support.
One is therefore facing the same localization problems of
eq.(2.1).

An advantage with respect to the Schroedinger equation is that the
minimal coupling electromagnetic interaction is described by local
terms, namely by the following equation \be{D_\mu\,D^\mu \ph(x) +
m^2 \ph(x) = 0, \,\,\,\,D_\mu \eqq \dmu + i e A_\mu(x).}\ee There
is however a serious  problem, namely the positive energy
condition is not stable under the interaction; in fact, quite
generally, even if the initial data satisfy the positive energy
condition, the Fourier transform of the corresponding solution may
contain negative frequencies.\footnote{ For example, an
interaction term $\U(x) \ph(x)$, in the Klein-Gordon equation,
with $\U(x)$ a potential of compact support in space and time,
induces transitions to negative frequencies since the frequency
spectrum of $\U$ is unbounded below. } This phenomenon is known as
the {\em Klein paradox} and represents a serious obstacle for the
interpretation of $\ph(\x, t)$ as the wave function of a quantum
particle.\footnote{ The analysis of the energy spectrum is
conveniently done in the (equivalent) first order formulation,
which in the free case reads $$ i\, \dt\, u = \left(
\begin{array}{cc} 0     &    \,\, 1 \\ -\Delta + m^2    &  \,\, 0
\end{array} \right ) u \eqq\,\, H_0 \,u, \,\,\,u =
\left(\begin{array}{c} u_1 \\ u_2
\end{array}\right) . $$ The Hilbert space $\H$ is defined by the
scalar product $$ (u, \,v) = \ume \int d^3 x\, [ \nabla u_1 \nabla
v_1 + m^2 u_1\,v_1 + u_2\,v_2].$$ Technically $u_1 \in
H^1(\Rbf^3), \,u_2 \in  L^2(\Rbf^3) $; the Hamiltonian $H_0$ is
selfadjoint on $D(H_0) = H^2(\Rbf^3) \oplus H^1(\Rbf^3)$ and its
spectrum is symmetric with respect to the origin. The interaction
with external (tempered) fields, e.g. the minimal coupling
electromagnetic interaction, corresponds to a bounded, in general
non symmetric, perturbation.}

In general the interaction with external fields having compact
support in time does not commute with the projection operator
$P_+$ which assures that the initial data (in the far past)
satisfy eq.(2.4) and therefore transitions to negative energies
are allowed.\footnote{For a discussion of the external field
problem see Ref. 3, in particular the contribution by R. Seiler.}

The roots of the above difficulties  are rather deep being related
to the impossibility of a non trivial time displacement of the
support of the initial data (corresponding to a non trivial
propagation speed) if the time evolution is described by an
evolution equation with
\newline i) {\em finite propagation speed (hyperboliticity)}
\newline ii) {\em localization of the solutions (compact support in space)}
\newline iii)  {\em positive energy spectrum}
\begin{Proposition} A time evolution with the above properties i)-iii)
cannot induce a  displacement of the support in space of the
initial data.
\end{Proposition}
\Pf  In fact, if $R$ is a (compact) region in space  disjoint from
the support $K$ of the initial data, then by hyperbolicity, for a
sufficiently small interval of time $I_t$, the solution remains
zero in $R$. Now, positivity of the energy spectrum implies that
the solution has an analytic continuation to imaginary times and
is analytic for $z = t - i \t, \,\,\,t \in \Rbf, \,\,\,\t > 0$
\be{\ph(\x, t - i \t) = (2 \pi)^{-1/2} \int d \o\,
\tilde{\ph}(\x,\o) \,e^{-i \o(t - i \t)},}\ee where $\tilde{\ph}$
denotes the Fourier transform of $\ph$ with respect to time. Thus,
by Schwarz's reflection principle for analytic functions,
\be{\ph(f, t) \eqq \int d^3 x\,f(\x)\,\ph(\x, t) = 0,}\ee if $
supp\,f \subseteq R,\, t \in I_t$ implies $\ph(f, t) = 0$, for $
\forall t$, i.e. $supp\, \ph(\x, t) \subseteq K, \forall t$.

\vspace{2mm}One  might think that the problems of  the
relativistic wave equations discussed above originate because time
and space derivatives do not appear in a symmetric way and that
the second time derivative gives rise to both negative frequencies
and a non positive density $\rho(\x)$. These seem to have been the
motivations for the Dirac equation, which is linear in time and
space derivatives at the expense of a four component wave
function; in the free case it reads \be{[ - i \g^\mu \dmu + m \,]
\psi(x) = 0, }\ee where $\g^\mu , \,\mu = 0, 1, 2, 3  $ are $4
\times 4$ matrices satisfying $\g^\mu \,\g^\nu + \g^\nu\,\g^\mu =
2 g^{\mu\,\nu}, $ with $g^{\mu\,\nu}$ the Minkowski metric.

One of the advantages with respect to the Klein-Gordon equation is
that the Dirac charge density  $j_0(x) = \psi^*(x) \psi(x)$ is
positive definite (with no condition) and can be interpreted as a
probability density. Another important nice feature of the Dirac
equation is that its time evolution is hyperbolic and that the
initial data can be taken of compact support. This property
persists for a large class of (local) interactions, including the
electromagnetic minimal coupling in the physical radiation
gauge.$^{(4)}$

However, as anticipated by the above Proposition 2.2, the spectrum
of the Hamiltonian $ H_0 = -i \a^i \di + \g^0 m, \,\,\,\,\a^i \eqq
\g^0 \g^i$, is not positive and one faces the problem of
eliminating the negative frequency solutions as before. Again one
could use initial data satisfying a positive energy condition
\be{P_+ \psi = \psi, \,\,\,\,\,P_+ = ( - \Delta + m^2)^{-1/2}
[\g^0 \sqrt{-\Delta + m^2} - i \g^i\,\nabla_i + m ], }\ee but i)
such a projection operator is non local, so that the initial data
satisfying the above condition cannot have compact support, ii)
time dependent interactions, e.g. an electromagnetic interaction,
induce transitions between positive and negative energy solutions,
as for the Klein-Gordon equation discussed above ({\em Klein
paradox} \footnote { See e.g. Ref. 5 and for a comprehensive
treatment Ref. 6.}). \footnote{Associated with the occurrence of
negative energy solutions is the problematic interpretation of the
coordinate $\x$ as the position operator; in fact its time
derivative $\dot{x}_i(t)$ has eigenvalues $\pm 1$, corresponding
(in our units) to the velocity of light ({\em Zitterbewegung});
moreover, even in the free case $\ddot{x_i} \neq 0 $ . Thus, as in
the Klein-Gordon case, the position operator is obtained by taking
projections on the positive energy subspace $$ \x^{op} = P_+ \x \,
P_+ , $$ with the inevitable delocalization problems discussed
above.}

To cure the negative energy problem, Dirac proposed his "hole
theory", according to which all negative energy states are
occupied and  Pauli exclusion principle precludes any transition
to them. Clearly, this step represents a radical departure from
Schroedinger (few particle) quantum mechanics, since the actual
picture involves both the Dirac wave function and the Dirac sea of
occupied negative energy states, i.e. infinite degrees of freedom.
Interactions can induce transition from a negative to a positive
energy state, giving rise to  holes in the Dirac sea, which will
appear as particles of opposite charge; the net result of such
transitions is creation of  pairs consisting of a positive energy
particle and a hole. It is not difficult to recognize in this bold
Dirac idea the seeds of quantum field theory, where the field
operators contain both positive and negative frequencies and the
positive energy spectrum is a property of the states.

\vspace{3mm} The difficulties of a one particle interpretation of
relativistic wave equations discussed above indicate a conflict
between  relativity and Schroedinger (few particle) quantum
mechanics. The net conclusion is that the one (or few) particle
picture is unstable against interactions and that almost
inevitably one has to admit the possibility of  many particles
excitations, i.e. a relativistic quantum mechanics must involve
infinite degrees of freedom.

\sloppy
\section{RELATIVISTIC PARTICLE INTERACTIONS AND QUANTUM MECHANICS}
\fussy Another source of problems for a combination of
Schroedinger quantum mechanics and relativity is their different
description of particle interactions.  In Schroedinger QM, the
treatment of interaction between particles is  based on the
canonical (Hamiltonian) formalism  and on the Newtonian concept of
{\em force at a distance}, (typically described by an interaction
potential), which makes use of simultaneity and therefore cannot
be relativistically invariant. As a matter of fact, there are
serious obstructions, even at the classical level, for building up
a relativistic dynamics involving forces at a distance. The
natural concept of interaction compatible with relativity is that
of contact force or more generally of local interaction with a
dynamical medium or a field. One is then led to abandon the
Newtonian picture of few particle interactions and consider the
infinite degrees of freedom associated with the (dynamical) field
responsible of the interaction.

\subsection{Problems Of Relativistic Particle Interactions}
A relativistic dynamics of particles in terms of distance forces
meets the problem that interactions cannot be instantaneous (an
inevitable delay resulting from the finite propagation speed) and
that simultaneity is not a relativistically invariant concept
$^{(7)}$. In fact, in the case of $N$ particles of definite
masses, if $x^{(i)}(\t^{(i)})$, denotes  the world line of the
$i$-th particle, $\t^{(i)}$ the corresponding proper time and $
\dot{x}^{(i)}_\mu \eqq d x^{(i)}_\mu / d \t^{(i)}$ the four
velocity, Lorentz invariance implies $\dot{x}^{(i)}_\mu
\,\dot{x}^{(i)\,\mu} = 1$. Hence, the stability of this condition
under time evolution imposes the following constraint on the
accelerations \be{\dot{x}^{(i)}_\mu \,\ddot{x}^{(i)\,\mu} = 0}\ee
and therefore on the forces.

Now,  a space-time translation invariant force at a distance on a
particle depends on the relative positions (and possibly on the
velocities) of the other particles at the same time, i.e. on the
(free) Cauchy data at the given time, so that in general eq.(3.1)
will not be satisfied. For example, as discussed  by Wigner,
$^{(8)}$ a space-time reflection invariant central force between
the $i, j$ pair of particles is of the form \be{F^{i j}_\mu =
(x^{(i)}_\mu - x^{(j)}_\mu) f = - F^{ j i}_\mu, }\ee where $f$ is
a function of the invariants which can be constructed in terms of
the four vectors $x_\mu$ and the four velocities; it is clear that
in general the four vector $ x^{(i)}_\mu - x^{(j)}_\mu$ will not
be perpendicular to both tangents of the two world lines.

To cure this problem, van Dam and Wigner $^{(9)}$ proposed to use
{\em non local}\, "forces at a distance", such that the force
$F^{(i j)}$  on the $i$-th particle by the $j$-th particle depends
on {\em all} the points of the trajectory of the $j$-th particle,
which are spacelike with respect to $x^{(i)}$.\footnote{Similarly,
in Feynman and Wheeler theory of particle interactions  the force
$F^{(i j)}$ depends on the points of the $j$-th trajectory which
lie on the light cone centered at $x^{(i)}$; with such a choice,
the conservation laws of the particle energy momentum are not
satisfied.} In this way, however, the dynamical problem is  no
longer formulated in terms of a Cauchy problem for differential
equations and  becomes almost untractable, since it involves the
knowledge of part of the particle trajectories.

No interaction theorems for relativistic particle dynamics have
also been proved within the canonical (Hamiltonian) formalisms, on
which quantum mechanics crucially relies.$^{(11)}$ A substantial
part for the proof of such results is the formalization of the
property of relativistic invariance. A simple and natural
translation of Poincar\'e invariance is that the ten generators of
the Poincar\'e group (space-time translations and Lorentz
transformations) are realized by functions of the canonical
variables and that their Lie algebra is satisfied with the Lie
product $[\,,\,]$ given by Poisson brackets $^{(12)}$; if $H,
\,P_i, \,J_i,\, K_i, \,i= 1, 2, 3, $ denote the generators of time
translations, space translations, space rotations and pure Lorentz
transformations, respectively, they must satisfy $$ [ H, \, P_i ]
= 0, \;\;[ H, \, J_i ] = 0, \;\;[ H, \, K_i ] = - P_i,$$ $$
[P_i,\, P_k ] = 0, \; \;[ \, P_i, \,J_k\,] = \eps_{i k l} P_l,
\;[\,P_i,\, K_k\,] = - \d_{i k} H,$$ \be{ [\,J_i,\, J_k\,] =
\eps_{i k l} J_l, \;[\,J_i, \, K_k\,] = \eps_{i k l} K_k, \;
[\,K_i, \,K_k \,] = - \eps_{i k l} J_l.}\ee Then one has
\footnote{ H. Leutwyler, loc. cit.}
\begin{Proposition} If the particle coordinates (on the trajectories)
$q^{\a}_i$, $\a = 1, 2, ...N$, transform correctly under the
Poincare' transformations, i.e. \be{ [\,q^{\a}_i, \,P_k\,] = \d_{i
k }, \,\,\,\,\,\,[\,q^{\a}_i,\, J_k\,] = - \eps_{i k l } q^{\a}_l,
\,\,\,\,\,\,[\,q^{\a}_i,\, K_k\,] = q^{\a}_i \,[\,  q^{\a}_k,\,
H\,],}\ee (world line conditions) and the equations of motions are
not degenerate, i.e. \be{ det \frac{\partial^2 H}{
\partial p^\a_i\,\partial p^\b_k} \neq 0,}\ee then the particle
accelerations vanish \be{ [\,[\,q^{\a}_i,\, H], \, H\,] = 0.}\ee
\end{Proposition}
{\bf Remark}  The non degeneracy condition states that the
positions and velocities form a complete set of dynamical
variables, so that the transition to a Lagrangian is possible in
the standard way. The above equations (3.4) for the transformation
properties of the  coordinates correspond to the world-line
condition of Currie et al. \footnote{For a simple derivation see
the review by Currie and Jordan pp. 93-94. We briefly sketch the
argument. Let $X_i \eqq q_i(t), \,\, X'_i \eqq q'_i(t')$ denote
the coordinates of particle position on the trajectory at time $t$
and the corresponding ones in a Lorentz transformed frame.  We put
$t' = 0$ and consider an infinitesimal transformation so that
second order terms in the Lorentz boost parameter $\a_j$ are
neglected. Then one has $$X_i = X'_i -  \a_j t' = X'_i, \,\,\,\,t
= t' - \a_j X'_j = - \a_j\,X'_j,$$  $$ X_i = q_i(t = - \a_j X'_j)=
q_i(0) - \a_j X'_j \,v_i = q_i(0) - \a_j q_j(0)\,v_i, \,\,\,\,v_i
\eqq d q_i(t)/d t|_{t=0}, $$ $$\a_j [ q_i, \,K_j\,] = \d^{\a_j}
q_i(t) = - (q_i'(0) - q_i(0)) = - \a_j q_j(0) v_i.$$ }.

\vspace{2mm}As discussed by Ekstein $^{(13)}$ the point at the
basis of the argument is that in the relativistic case the
Hamiltonian can be obtained from the Lie product of $[\,P_i,\,
K_i\,]$ and therefore if space translations and boosts have a
kinematical character, i.e. can be written as sums of single
particle functions, so is also the Hamiltonian and there cannot be
any particle interaction. Such a constraint does not exists in the
non relativistic case, where the non vanishing Lie products of the
generators of the Galilei group are $[\,G_i, \,H\,] = P_i$, $
[\,P_i, \, G_k\,] = m \d_{ i\,k}$ and those which state the vector
character  of $P_k, \,J_k,$ and $G_k$ (the generators of the
accelerations) and . As it is well known, such  Lie products are
compatible with a non trivial particle interaction.

\def \A  {{\cal A}}
\sloppy
\subsection{Field  Interactions And Quantum Mechanics}
\fussy
The difficulties pointed out above for interactions
described by distance forces suggest to consider contact forces or
more interestingly field mediated interactions, with a field
contact action on the particles. In this case, the interaction is
a result of energy momentum exchanges between the particles
through the field, which propagates energy and momentum and can
transfer them to the particles by contact. Lorentz covariance is
then transcribed in the Lorentz invariance of the field equations.
This appears therefore as the distinguished (if not the exclusive)
way of formulating relativistic  particle interactions and indeed
the electromagnetic interaction is the prototypical  example.

When such a picture is confronted with quantum mechanics,
interesting considerations emerge, as clearly emphasized by
Heisenberg $^{(14)}$.  If the classical particles are promoted to
Schroedinger particles, according to the principles of quantum
mechanics, the question arises about the quantum mechanical status
of the interaction mediating field. The possibility of keeping a
classical structure for the fields is ruled out by Heisenberg
uncertainty relations.

The point in Heisenberg argument is that if the measurement of
field momentum and its localization were not constrained by
quantum mechanical limitations, one could use the particle-field
interaction to violate the Heisenberg uncertainty relations for
the measurement of the particle position and momentum.

A relativistic description of particle interactions mediated by a
local action of fields on the particles  poses the problem of the
field dynamics, which, as argued before, should be described by
{\em local relativistic field equations}. We have already seen
such equations in Sect.\,2, albeit with different motivations, and
one would run into the same problems discussed before if the
fields are interpreted as Schroedinger wave functions. In
particular, positive energy spectrum would exclude localizability
of the field. Again a departure from Schroedinger quantum
mechanics in the direction of field quantization offers a solution
of such a problem. In fact, the classical expression of the field
energy, e.g. in the free case, \be{ H(\ph) = \ume \int d^3 x \,[
\nabla \ph^*\,\nabla \ph + m^2 \ph^*\,\ph +
\dot{\ph}^*\,\dot{\ph}\,]}\ee is positive definite and, in the
canonical formulation, the Poisson brackets of $H(\ph)$ give the
time derivatives of the canonical field \be{  \dt\, \ph = - \{
H(\ph), \,  \ph\,\}.}\ee Thus, one recovers the r\^{o}le of
$H(\ph)$ as the {\em  positive} quantum generator of time
translations, provided the field is considered as a quantum
operator ({\em field quantization}) and the Poisson brackets are
replaced by commutators, as in the canonical quantization
procedure of classical theories. In this way, {\em positivity of
the energy} is obtained. Furthermore, with  the promotion of the
field from Schreodinger wave function to a quantum operator, the
frequency spectral support of  (the Fourier transform of) the
field operator does not describe the energy-momentum of a state
(as it would be the case for a wave function) but rather the
energy-momentum {\em variations} that the field operator may
induce by its application on a state, so that it is not
constrained to lie in the forward cone $\bar{V}_+$ by the spectral
condition on the energy momentum operator. Then, a {\em local
field dynamics} is allowed.

In this picture,  also the problems connected with the non
positive definiteness of $j_0(x,t)$ disappear, since $j_0$ has the
meaning of charge density operator, whereas the probability
density, say  for a one particle state $\Psi$, is given by the
modulus squared of its (c-number) wave function $\Psi(\x,t)$,
which is related to (suitable) matrix elements of the quantum
field operator $\ph(\x, t)$.

As it is well known and it is instructive to check in the above
perspective, all this works very well in the free case. In
conclusion, the problems of local relativistic equations of motion
and positive energy spectrum are solved in the following way: i)
the local field equations allow for  both positive and negative
frequencies and in fact the field has support in both the upper
and lower hyperboloids $p^2 = m^2$, ii) the Hamiltonian operator
is a positive operator and therefore its spectrum is positive,
iii) the negative frequency part of the field, say $\ph^-$, is an
operator whose action on a state lowers its energy; the positivity
of the energy spectrum then requires that there is a lowest energy
energy state (vacuum state) which is annihilated by $\ph^-$.

All this strongly suggests that relativistic particle interactions
are accounted for by the absorption or emission of energy momentum
quanta carried by the interaction mediating field. The question of
whether such an idea leads to a mathematically consistent theory
of particle interactions is the fundamental problem raised in the
Introduction.
\subsection{General Properties Of Quantum Field Theory}
The investigation of the non perturbative foundations of
relativistic quantum field theory began soon after the success of
the perturbative expansion of quantum electrodynamics, when the
convergence of the series was seriously questioned, the theory was
predicted to be afflicted by ghosts ${(15)}$ and the high energy
behaviour seemed to indicate the inconsistency of the theory.

Furthermore, the mathematical problems  of the interaction picture
codified by Haag theorem and by Powers and Baumann theorems on the
impossibility of using canonical quantization for the quantization
of interacting fields, indicate that a non perturbative foundation
of quantum field theory is an unavoidable issue.

The two major steps in this direction were the Wightman
formulation of quantum field theory in terms of the so called
Wightman "axioms", $^{(16)}$ which emphasize the spectral
condition, relativistic invariance and locality, and the Lehmann,
Symanzik and Zimmermann approach (LSZ "axioms") $^{(17)}$  based
on the asymptotic condition and the $S$-matrix elements. We shall
briefly discuss the first approach, which proved to be more
powerful, the second being derivable from it $^{(18)}$ . The term
"axioms", even if correctly stressing the attention to the
mathematical rigor  and consistency, does not do justice to the
fact that in both cases they represent the mathematical
formulation of deep and simple physical requirements.

The quantum mechanical interpretation of the theory and stability
are encoded in the following {\bf quantum mechanical properties}
\vspace{1mm} \newline {\bf QM1.} ({\bf Hilbert space structure})
The states are described by vectors of a (separable) Hilbert space
$\H$ \vspace{1mm}
\newline {\bf QM2.} ({\bf Energy-momentum spectral condition}) The
space-time translations are a symmetry of the theory and are
therefore described by strongly continuous unitary operators
$U(a), \, a \in \Rbf^4$, in $\H$.

The spectrum of the generators $P_\mu$ is contained in the closed
{\em forward cone} $\bar{V}_+ = \{p_\mu: p^2 \geq 0, \,p_0 \geq 0
\}$. There is a {\em vacuum state} $\Psio$, with the property of
being the unique translationally invariant state, (hereafter
referred to as  {\em uniqueness of the vacuum}). \vspace{1mm}
\newline {\bf QM3.} ({\bf Field operators}) The theory is formulated in terms of
fields $\ph_k(x)$, $k = 1, ...N$, which are operator valued
tempered distributions in $\H$, with $\Psio$ a {\em cyclic} vector
for the fields, i.e. by applying polynomials of the (smeared)
fields to the vacuum one gets a dense set $\D_0$.

\vspace{3mm} {\bf Remarks}. The separability of $\H$ actually
follows from the cyclicity of the vacuum and temperedness, since
$\S(\Rbf^4)$ has a countable basis \footnote{Both the test
function spaces $\S$ and $\D$ are separable as topological
spaces}.

The cyclicity of the vacuum states that the  fields provide a
complete set of "dynamical variables" in terms of which, by
application to the vacuum, one can describe all the states of
$\H$.

\vspace{2mm} The relativistic invariance of the theory  is
formalized by the  following {\bf relativistic properties}:
\vspace{1mm} \newline {\bf R1.} ({\bf Relativistic covariance})
The Lorentz transformations $\Lambda$ are described by (strongly
continuous) unitary operators $U(\Lambda(A))$, $ A \in  SL(2,
\Cbf)$ = the universal covering group \footnote{This accounts for
the spinor representations; for simplicity, in the following the
"label" $A$ in $\La(A)$ will not be always  spelled out.} of the
restricted Lorentz group $L_+^\uparrow$ (characterized by $det\,
\Lambda = 1, \,sign \,\Lambda_0^0 > 0$) and the fields transform
covariantly under the Poincar\'e transformations $U(a, \,\Lambda)
= U(a)\,U(\La)$: \be{U(a, \La(A)) \,\ph_i(x)\,U(a, \La(A))^{-1} =
S_{i\,j}(A^{-1})\,\ph_j(\La x + a),}\ee with $S$ a finite
dimensional representation of $SL(2, \Cbf)$.
\vspace{1mm}\newline{\bf R2.} ({\bf Microcausality or locality})
The fields either commute or anticommute at spacelike separated
points \be{ [ \, \ph_i(x), \,\ph_j(y)\,]_{\mp} = 0,
\,\,\,\mbox{for}\,\,\,(x - y )^2 < 0.}\ee

\vspace{3mm} {\bf Remarks}. Eq. (3.9) guarantees the manifest
covariance of the formulation, but is not strictly implied by
relativistic invariance, since one could use non covariant fields;
in this sense it is an essentially technical condition, since it
is clearly more convenient to use dynamical variables with simple
transformation properties under the symmetries of the theory.

It is easy to see  that eq.(3.10) holds for free fields; in the
interacting case its validity for observable fields, like e.g. the
electromagnetic field $F_{\mu\,\nu}(x)$, is required by Einstein
causality, by which measurements of observables localized in
spacelike separated regions must be always compatible (in the
quantum mechanical sense) and therefore such observables must
commute. The validity of eq.(3.10) for unobservable fields, like
e.g. the fer\-mion fields (or fields carrying  a gauge charge), is
an extrapolation with respect to the physical requirements, on the
basis of the free case and should be considered as an essentially
technical requirement, like the field manifest covariance. The
main motivation is that it guarantees that the observable
operators constructed in terms of unobservable local fields, like
e.g. the current density, the momentum density etc., automatically
commute at spacelike separated points, i.e. satisfy Einstein
causality. It is also fair to say that most of our present wisdom
on QFT, including the full control of the low dimensional cases,
comes from models formulated in terms of local (in general
unobservable) fields.

An alternative approach is to use only (bounded) local
observables; in this case not all the physical states belong to
the vacuum representation of the observable algebra and
unobservable fields may possibly be constructed as intertwiners
between inequivalent representations of the observable algebra
$^{(19)}$ . This approach, also called algebraic quantum field
theory, is more economical from a conceptual point of view, but
less practical  for the constructive problem of quantum field
theory models. One of the main virtues of Wightman approach is its
strong link with the conventional wisdom of quantum field theory,
including the perturbation theory, which crucially relies on the
use of non observable fields, like the vector potential and the
electron field in QED. For this reason, the investigation of the
mathematical structures of the  vacuum expectation values of
fields, which are at the basis of Wightman approach, has  a direct
impact on the conventional formulation.

The properties QM1-QM3,  R1, R2 imply the following properties of
the vacuum expectation values ({\em Wightman functions}) of, e.g.,
a scalar field. As a consequence of QM3 \vspace{1mm} \newline {\bf
W1.} $\W(x_1, ...x_n) \eqq ( \Psio, \ph(x_1)...\ph(x_n)\,\Psio )$
are tempered distributions.

\vspace{2mm}In the following for brevity we shall use a
multivector notation $\W(x) = \W(x_1, ...x_n), \, x = (x_1,
...x_n).$

\def \tW {\tilde{W}}
\def \bV {\overline{V}_+}
As a consequence of QM2 and R1, putting $\xi_j \eqq x_{j+1} -
x_{j}$, one has  \vspace{1mm} \newline {\bf W2.} ({\bf
Covariance}) \be {\W(x_1, ...x_n) = W(\xi_1, ...\xi_{n-1}) \eqq
W(\xi) = W(\La \xi).}\ee {\bf W3.} ({\bf Spectral condition}) The
support of the Fourier transform $\tW$ of $W$ is contained in the
product of forward cones,   i.e. \be{ \tW(q_1, ...q_{n}) =
0,\,\,\,\,\,\mbox{if}\,\,\,\, q_j \notin \bV.}\ee
\vspace{1mm}{\bf
W4.} ({\bf Locality}) R2 gives \be{ \W(x_1, ...x_j, x_{j+1},
...x_n) = \W(x_1, ...x_{j+1}, x_j, ...x_n),
\,\,\,\,\mbox{if}\,\,\,\,(x_j - x_{j+1})^2 < 0.}\ee

The Hilbert space structure of QM1  gives \vspace{1mm}\newline{\bf
W5.} ({\bf Positivity}) For any terminating sequence $
\underline{f} = (f_0, f_1, ...f_N)$, $f_j \in \S(\Rbf^4)^j$ one
has \footnote{This is the transcription of positivity of the norm
of any state of the form $$\Psi_{\underline{f}} = f_0 \Psio +
\ph(f_1)\,\Psio + \ph(f_2^{(1)})\,\ph(f_2^{(2)})\Psio +...,
\,\,\,$$ where $\underline{f} = (f_0, \,f_1,... f_N)$, $\,f_j =
\prod_{k = 1}^j f_j^{(k)}(x_k)$.} \be{ \sum_{j,k}\int d x \,d y
\,\bar{f}_j(x_j, ...x_1)\,f_k(y_1, ...y_k)\,\W(x_1,...x_j; y_1,
...y_k))\, \geq 0.}\ee

The uniqueness of the translationally invariant state is
equivalent to \vspace{1mm}\newline {\bf W6}. ({\bf Cluster
property}) For any spacelike vector $a$ and for $\l \ra \infty$
\be{ \W(x_1, ...x_j, x_{j+1} + \l a, ...x_n + \l a) \ra \W(x_1,
...x_j)\,\W(x_{j+1}, ...x_n),}\ee (the convergence  being in the
distributional sense).

\vspace{1mm}Eq.(3.15) says that the correlation function of two
mo\-no\-mials of (smea\-red) fields,  called clusters, factorizes
in the limit of infinite spacelike distance between the two
clusters. This property plays a crucial r\^ole for the existence
of asymptotic (free) fields and therefore for the construction of
the $S$-matrix, as clarified by the Haag-Ruelle theory $^{(20)}$.
It corresponds to a sufficient fall off of the potential in
potential scattering and it is one of the basic axioms of the
so-called $S$-matrix theory $^{(21)}$. Property (3.15) is related
to the independence of events associated to two clusters, when
their separation becomes infinite in a spacelike direction; the
rate by which the limit is reached in eq.(3.15) has actually been
related to the decay rate of the "force" between the two clusters
$^{(22)}$

\vspace{1mm} From the general physical requirements, transcribed
in the mathematical properties W1-W6, a number of relevant
structural information have been derived, some of them having
direct experimental consequences.

The first point is that the conditions W1-W6 provide a more
general quantization rule than canonical quantization. Actually,
in the case of a field  obeying a free field equation, W1-W6 imply
canonical quantization as a result of  Jost-Schroer theorem,
\footnote{R.F. Streater and A.S. Wightman, loc. cit. Theor.4-15.}
whereas for a general class of interacting fields, as proved by
Powers and Baumann, W1-W6 exclude canonical quantization and
provide a (non perturbative) substitute for it.

Another important consequence of W1-W6 is the existence of
asym\-ptotic fields and, under the assumption of asymptotic
completeness, of a unitary $S$-matrix. Also the LSZ asymptotic
condition and the associated reduction formulas can be derived
from the Wightman formulation, as proved by Hepp (see his Brandeis
lectures); furthermore one can prove dispersion relations for
scattering amplitudes, yielding experimentally measurable
relations $^{(23)}$.

By exploiting the analyticity properties of the Wightman
functions, a proof of Pauli principle has been obtained, namely
that in the alternative of commutation or anticommutation
relations at spacelike points, fields carrying (half)integer spin
must (anti)commute (this is the famous Spin Statistic Theorem).

The validity of the $P C T$ symmetry, (up to now an experimentally
established property) also follows from the general properties of
the Wightman functions, being in particular related to local
commutativity (see the Streater and Wightman book).

Finally, by exploiting the spectral condition, the Lorentz
covariance and locality,  it has been shown that the Wightman
functions have an analytic continuation to the so called euclidean
points and therefore in this way one derives the existence and the
general properties of euclidean quantum field theory, a crucial
technical tool for the non perturbative approaches developed in
the last decades (like the lattice approach to gauge theories, the
constructive strategy etc.).

Unfortunately, the physically motivated constraints W1-W6 are
highly non trivial to satisfy, as indicated by the non
perturbative results on the triviality of the $\l \ph^4$ theory
and possibly of quantum electrodynamics. It may be instructive to
note that the $\l\,\ph^4$ theory in four space time dimensions
would no longer be trivial if either the Hilbert space structure
(i.e. positivity) or the spectral condition ($\l > 0$) is relaxed
$^{(24)}$.

\section{QUANTUM GAUGE FIELD THEORY}
A widespread belief is that the above mentioned triviality
problems do not exist for asymptotically free QFT's and in
particular for a large class of non abelian gauge theories. Thus,
at present, gauge field theory appears as the good candidate for
the solution of the problem of combining quantum mechanics and
relativity.

 Unfortunately, in gauge  theories the (positive Hilbert space)
quantization of charged fields requires their strong
delocalization, so that locality, eq.(3.10), fails. Since all the
wisdom gained from perturbation theory crucially relies on the use
of local fields and very little is known about non local field
theories, a relevant mathematical question is which are the
alternatives and what can be learned from the perturbative
approach about the general structures. This is the content of the
present section.

From the standard perturbative treatment of gauge theories one may
be led to believe that the differences with respect to the
ordinary QFT models, like $\l \ph^4 $, are merely technical and
essentially amount to  more complicated bookkeeping rules and to a
proliferation of indices. In fact, in quantum field theory
textbooks, the perturbative expansion of QED in the Feynman gauge
is discussed in strict analogy with the scalar field theory model.

In effect, even from a physical point of view, gauge field
theories are characterized by radically different structures with
respect to ordinary QFT's and such differences are responsible for
the new problems which arise in the quantization of gauge
theories.

The basic difference is the invariance under the infinite
dimensional Lie group $\G$ of (local) gauge transformations, so
that by the second Noether theorem one not only deduces the
existence of a conserved current $j_\mu^\a$, for each generator of
the subgroup $G$ of rigid (or global) gauge transformations, but
also the fact that such a current is the divergence of an
antisymmetric tensor (in the language of differential geometry the
associated form is a boundary) \be{ j_\mu^\a = \partial ^\nu
G^\a_{\nu\,\mu}, \,\,\,G^\a_{\nu\, \mu} = - G^\a_{\mu \, \nu},
}\ee $ \a = 1, ...N$, $N = dim \,G$. Such a property will be
called (local) {\bf Gauss law}. It plays a rather crucial role for
understanding the structural differences with respect to ordinary
QFT's , since most of the peculiar phenomena exhibited by gauge
field theories, like Higgs mechanism, linearly raising quark-anti
quark potential, quark confinement, charged field delocalization
etc. depend on it.

We briefly mention some of its important consequences
\footnote{For a look to gauge theories which emphasizes the local
Gauss law, see Refs. 25.}. First, current conservation becomes a
geometrical identity which does not involve the evolution
equations of the charge carrying fields. Secondly, the Gauss' law
implies the superselection of the charge associated to the current
$j_\mu^\a$, formally \be{Q^\a = \int d^3 x\, j_0^\a(\x, t),}\ee
i.e. one not only has a selection rule associated to the
conservation law $[\,Q^\a,\, H\,] = 0$, with $H$ the Hamiltonian,
but the much stronger property that $[\,Q^\a, \,A\,]$, for {\em
any} (local) observable $^{(26)}$ .

Finally, the  Gauss' law implies the non locality of the charged
fields $^{(27)}$, since,  e.g. in the abelian case, a field $\ph$
has a charge $q$ if \be{[\, Q, \,\ph\,] = q \ph,}\ee and by Gauss'
law one has \be{ [\,Q, \,\ph(y)\,] = \int d^3 x \,[\, j_0(\x, 0),
\, \ph(y)\,] = [\, \Phi_\infty({\bf E}), \, \ph(y)\,], }\ee where
$\Phi_\infty({\bf E})$ denotes the flux at spacelike infinity of
the electric field $E_i \eqq G_{0\,i}$. Clearly, if $\ph(y)$ and
$E_i(x)$ commute at spacelike separations, the right hand side
vanishes (since the spacelike infinity is spacelike with respect
to any spacetime point $y$) and $\ph(y)$ cannot have a non zero
charge $Q$. The physical meaning of such a property is rather
transparent, being related to Gauss' theorem; the local Gauss' law
implies long range correlations and a Coulomb like delocalization
to ensure a non vanishing electric flux at infinity.

In the standard perturbative approach, e.g. in the so called
covariant and renormalizable gauges, the non locality implied by
Gauss' law is somewhat hidden by the choice of weakening the
requirement of gauge invariance and by requiring the Gauss' law
only on a subspace of the vector space of the corresponding
quantum field theory (see below). This choice leads to the
violation of positivity by the correlation functions of the
charged fields.

In conclusion, quite generally one can prove that in the
quantization of gauge field theories the (correlation functions of
the) charged fields cannot satisfy all the quantum mechanical
constraints QM1, QM2 and the relativity constraints R1, R2, since
locality and positivity are crucially in conflict $^{(28)}$.
Therefore, the general framework discussed in Sect.3.3 has to be
modified ({\em modified Wightman axioms}).
\subsection{Quantum Mechanics And Gauss' Law}
To better grasp the structural problems arising in the
quantization of gauge theories the following brief digression
about charged states may be of help.

In standard {\bf QFT models with no superselection rule}, i.e.
with no quantum number commuting with all the observables, the
vacuum is a cyclic vector for the (local) observable field algebra
$\A$, in a Hilbert space $\H$, which contains  all the physical
states (which can be associated to such a vacuum state).

In general, a {\bf gauge QFT} is characterized by the existence of
quantum numbers $Q^\a$, called "charges", which commute with all
the observables; the observable (local) field algebra $\A$ is a
proper subalgebra of a field algebra $\F$ and relevant particle
representations of $\A$,  defined by "charged states", are
disjoint from the vacuum sector \be{\H_0 \eqq \overline{\{\A
\Psio\}}.}\ee The Hilbert space $\H$ of physical states decomposes
as a direct sum of spaces $\H_q$ of definite charge $q$, called
charged sectors, and the charged fields intertwine between the
vacuum and the charged sectors \footnote{In the mathematical
language they define morphisms $\rho_q$ of the (local) observable
algebra, so that a charged state representation $\pi_q$ can be
obtained from the vacuum representation $\pi_0$ by a morphism
$\rho_q$: $\pi_q(\A) = \pi_0(\rho_q(\A))$. For an extensive
analysis of the construction of charged states in terms of
morphisms of the observable algebra see R. Haag, {\em Local
Quantum Physics}, Springer 1996, esp.Ch.\,IV.}. A typical example
of this structure is provided by quantum electrodynamics in the
Coulomb gauge, where the electron field connects the vacuum sector
to the charged sectors.

\vspace{2mm}In the class of gauge theories, one can furthermore
distinguish two cases.

\vspace{1mm} A. {\bf Gauge QFT of DHR type}$^{(29)}$ . There is a
global gauge group $G$ (a finite dimensional compact Lie group)
leaving the local observables pointwise invariant, with no Gauss'
law associated to the corresponding superselection rules. In this
case, the charged sectors are described by {\em local fields} or
more precisely by {\em local morphisms}; an example of this type
would be given by a Yukawa theory of nucleon-pion interactions, in
the case in which all the observables are invariant under the
$SU(2)$ isospin group.

\vspace{2mm} B. {\bf Gauge QFT with Gauss' law}. The local
observables are pointwise invariant  under  a local gauge group
(an infinitely dimensional Lie group), equivalently a local Gauss
law holds. In this case the charged sectors cannot be described by
local fields or by local morphisms and the DHR classification and
theory does not apply. A typical example is given by QED.
\footnote{For such a theory the delocalization of charged sectors
is stronger than expected on the basis of classical
electrodynamics. In fact, whereas one expects a  delocalized
Coulomb tail for the expectations of the electric field on a
charged state, one would expect localization for the charge
density $j_0(x)$ on charged states; actually, it is not so (D.
Buchholz, S. Doplicher, G. Morchio, J. Roberts and F. Strocchi,
Ann. Phys. {\bf 290}, 53 (2001)).}
\subsection{Renormalizable Gauges And Quantum Mechanics}
Since, the perturbative expansion of QED has been developped in
the so called renormalizable gauges, in particular in the Feynman
gauge, it is perhaps interesting to discuss the mathematical
properties of such a gauge and its quantum mechanical
interpretation.

In the Feynman quantization of QED, the field algebra is generated
by the {\em (local) fields} $\psi, \, \overline{\psi}$, and
$A_\mu$, and their vacuum correlation functions satisfy the
relativistic spectral condition and Lorentz covariance. The field
equations \be{ j_\mu = \Box A_\mu
=
\partial^\nu F_{\nu\, \mu} + \dm \partial A}\ee display the violation
of Gauss' law, by the occurrence of the longitudinal field $ \dm
\partial A $, which cannot vanish if the
electron fields are local ({\em weak Gauss'  law}). Quite
generally, one can show that the correlation functions of the
fields cannot satisfy positivity and therefore there is no obvious
Hilbert space structure associated to them.

The solution of this problem in the free case is well know (and
discussed in any textbook on QED) and essentially due to Gupta and
Bleuler ({\em Feynman-Gupta-Bleuler quantization}). The lesson for
the general case is that the vacuum expectations $< \,\,>_0 $ of
the fields define a vector space $\V = \F \,\Psio$, and  an inner
product on it by \be{< A \,\Psio, \, B\,\Psio  > \eqq < A^*\, B
>_0, \,\,\,\forall \,A, \, B \in \F .}\ee The subspace $\H' \subset \V$,
characterized by the property of yielding vanishing expectation
\footnote{Such a non linear condition is replaced by a linear {\em
supplementary condition}, which in the abelian case reads
$(\partial A)^- \H' =0,$ with the minus denoting the destruction
part of the free field operator $\partial A$ and in the non
abelian case reads $Q_{BRST} \H' = 0, $ where $Q_{BRST}$ is the
BRST charge.} of the longitudinal field $ \dm
\partial A $, is a candidate for the physical space. In fact, the
validity of the Gauss' law in expectations on $\H'$ corresponds to
the gauge invariance condition of the physical states.
Furthermore, the inner product $< .,\, .>_0$ is non negative on
$\H'$, so that $\H'$ has a pre-Hilbert structure and a quantum
mechanical interpretation is possible.

General conditions assure that the inner product $< . ,\, . >$ can
be  related to a Hilbert product $ ( .,\, . )$ on $\V$, by a
metric operator $\eta$, with $\eta^2 = 1$, \be{ < A, \, B > = ( A,
\,\eta B )}\ee ({\em Hilbert-Krein structure}). The construction
of $\eta$ is linked to the infrared problem and physical charged
states belong to the Hilbert-Krein closure of $\V$, i.e. can be
obtained as weak limits of (local) vectors of $\V$; in this way,
one obtains a solution of Zwanziger problem \footnote{For a more
detailed discussion of these structures arising in the
Feynman-Gupta-Bleuler formulation of QED, see Ref. 30.}

\subsection{Locality, Gauss' Law and Non Regularity}
In the Feynman gauge the quantization of quantum electrodynamics
is obtained by adding to the gauge invariant Lagrangean a so
called gauge fixing, which breaks the invariance under the full
gauge group down to the subgroup of gauge transformations with
gauge parameter $\Lambda(x)$ satisfying the free wave equation.
Such a breaking of gauge invariance leads to the weak form of the
Gauss' law. An alternative strategy, pioneered in particular  by
Wilson for gauge theories on a lattice, is to use a fully gauge
invariant Lagrangean in the euclidean action, i.e. to avoid the
introduction of   a gauge fixing. In this way, the so obtained
correlation functions satisfy locality, spectral condition and
positivity and the Gauss' law holds in operator form. However, in
such an approach to the quantization of gauge theories, the
representation of the charged fields is rather peculiar,  since
all their correlation functions (at non coincident points), in
particular the charged field propagators, vanish. This excludes a
connection with the standard perturbative approach and it is
actually incompatible with canonical quantization, since the so
represented charged fields cannot obey canonical commutation
relations.

Furthermore, in such an approach to the quantization of gauge
theories,  the Hilbert space defined by the so obtained
correlation functions consists of the vacuum sector and the access
to the charged states representations is not trivial.

\vspace{2mm}A further alternative, with the purpose of keeping
locality, positivity and the positivity of the energy, is the {\em
temporal gauge quantization} \footnote{For a review of the
literature on such a quantization, see Ref. 31.}.

The incompatibility of the Gauss'  law with the canonical
commutation relations of the vector potential $A_i, A_0 = 0$,
which is clearly displayed in the abelian  case by the equation
\be{ [\,(div E - j_0)(y), \,\,div A(x)\,] = - i \Delta \d(y -
x),}\ee is at the basis of the contradictory proposals, discussed
in the literature, for the propagator of the vector potential,
with peculiar features, like the breaking of time translations,
characterizing the various solutions.

A clarification of the mathematical structure of the temporal
gauge in the QED case and its compatibility with the basic quantum
mechanical requirements has been discussed recently $^{(32)}$. In
particular, one can show that the Gauss' law in operator form
requires a {\em non regular representation} for the vector
potential.

To make this point clear, it is useful to distinguish, in the
mathematical structure of quantum mechanics, the algebraic
properties of the canonical variables, i.e. the algebra of the
canonical commutation relations (briefly the CCR algebra) and its
representation in terms of Hilbert space operators.

As  CCR algebra it is convenient to take the algebra $\A_W$
generated by  the Weyl operators $U(\a) \sim\,e^{ i \a q}$, $V(\b)
\sim \,e^{i \b p}$. The use of bounded functions of the $q$'s and
$p$'s, avoids domain questions and allows to give $\A_W$ a $C^*$
algebraic structure.

A representation of the CCR algebra $\A_W$ in terms of Hilbert
space operators is called {\em regular} if the (representatives of
the) Weyl operators are strongly continuous in the parameters $\a,
\,\b$. This property characterizes the standard Schroedinger
quantum mechanics (of systems with finite degrees of freedom),
since the $q$'s and $ p$'s can be obtained by derivation of the
Weyl operators and by the Stone-Von Neumann theorem, apart from
multiplicities, there is only one regular representation. For
these reasons, in the textbook presentations of the basic
structure of quantum mechanics the above distinction is usually
omitted.

Non regular representations have been dismissed as pathological in
the past, but they are actually unavoidable for interesting
physical problems, also in the case of systems with finite degrees
of freedom, like the Bloch electrons and the associated $\theta$
states, the ground state representation of a quantum particle in a
periodic potential, the quantum particle on a circle etc.
\footnote{The general features of the quantum mechanics arising
from non regular representations have been discussed in Ref. 33. A
generalization of the Stone-Von Neumann theorem yielding a
classification of the non regular representations of the CCR
algebra, which are strongly measurable, has been given in Ref. 34.
 For the discussion of the quantum mechanical
examples see Ref. 35.} A typical example of a non regular
representation is given by the state defined by the following
expectations \be{ < U(\a)\,V(\b) > = 0, \,\,\,\mbox{if}\,\,\,\,\a
\neq 0, \,\,\,\,< V(\b) > = 1.}\ee In this representation the
position operator cannot be defined (only the corresponing Weyl
operator $U(\a)$ exists) and in fact, the common physical feature
of the above mentioned quantum mechanical examples is that the
ground state (wave function) is so delocalized, (as a consequence
of periodic structures), that it is impossible to define the
position. Non regular representation also appear in QFT models in
connection with infrared singular (canonical) fields, as in the
algebraic fermion bosonization, in two dimensional conformal
models,   in the positive quantization of the massless scalar
field in two dimensions, in the positive gauge quantization of the
Schwinger model, in  Chern-Simons quantum field theory models
$^{(36)}$. Also in these QFT examples, the occurrence of an
infrared delocalization is responsible for the non regularity.

For the temporal gauge, its realization without violating
locality, Hilbert space structure and operator Gauss' law is
characterized by the following vacuum expectation, for any test
function $h$, \be{ < e^{ i div A(h)}
>_0 = 0, \,\,\,\,\mbox{if}\,\,\, h \neq 0, }\ee where $div A(h) = \int
d^4x\, div A(x)\,h(x)$. This means that the longitudinal algebra
cannot be regularly represented and the correlation functions of
the vector potential, in particular its propagator, do not exits
(only those of its exponentials do).

In conclusion, the quantization of the temporal gauge requires a
departure from the standard quantum mechanical structure, by
allowing for a non regular representation of the longitudinal
field algebra. \footnote{A quantization of the temporal gauge in
which the two point function of the vector potential exists and
satisfies locality, positivity of the energy spectrum, invariance
under space time translations, rotations and parity, requires  a
weak form of the Gauss' law and a violation of positivity; see
Ref. 32.}
\subsection{Non Local Charged Fields}
Finally, we shall briefly mention the basic features of
quantizations of gauge theories, which keep the Gauss' law in
operator form, the Hilbert space structure or positivity and allow
for {\em non local charged fields}. This is the case of the
Coulomb gauge quantization of QED.

On one side, one gets most of the nice features of a standard
quantum field theory; on the other side, as mentioned before, the
non locality of the charged fields is the origin of technical and
conceptual problems.

\vspace{2mm}The non local dynamics of the charged fields, with
typical Coulomb delocalization, gives rise to  a crucial
dependence on the boundary terms in finite volume and on variables
at infinity in the termodynamical limit, with phenomena common to
gauge theories and Coulomb systems $^{(37)}$.

In a non perturbative constructive approach, the removal of the
infrared cutoff and the renormalization is more difficult.

Particles states are described by non local fields and the
standard proofs of PCT  and Spin Statistics theorems do not apply.

The relation between the charge density and the electric charge
becomes more delicate $^{(38)}$.

\vspace{3mm}One may be led to think that most of the problems
discussed above for the quantization of gauge theories are gauge
artifacts and one should not take them seriously. As a matter of
fact, they are not merely technical problems with little or even
no physical relevance. They hinge on the  theoretical problem of a
non trivial theory which combines quantum mechanics and relativity
and reflect the physical crucial fact that gauge invariance (or
equivalently Gauss' law) gives rise to a strong delocalization of
the (physical) charged states, so that they cannot be described by
fields which satisfy local commutativity.

\vspace{10mm}

REFERENCES \vspace{5mm}\newline1. R. Fernandez, J. Frohlich and
A.D. Sokal, {\em Random Walks, Critical Phenomena, and Triviality
in Quantum Field Theory}, Springer 1992, esp. Sects. 1.5.1.5-6,
1.5.2.
\newline 2. F. Dyson, {\em Phys. Rev.} {\bf 85}, 631 (1952)
\newline 3. G. Velo and A.S. Wightman eds.,  Proceedings of the 1977 Erice School, {\em
Invariant Wave Equations},  Springer 1978.
\newline 4.  D. Buchholz, S. Doplicher, G. Morchio. J. Roberts and
F. Strocchi, {\em Ann. Phys.} {\bf 209}, 53 (2001), Sect. 2.
\newline 5. B. Thaller, {\em The Dirac equation}, Springer
1992, Sect. 4.5 and p. 307.
\newline 6. A.S. Wightman, "Invariant
wave equations; general theory and applications to the external
field problem," in {\em Invariant Wave Equations}, Erice 1977, G.
Velo and A.S. Wightman eds., Springer 1978, pp.76ff .
\newline 7.  L.D. Landau and E. Lifshitz, {\em The classical theory of
fields}, Addison-Wesley 1962, Ch.III, Sect.15.
\newline 8. E.P. Wigner, "Relatistic Interaction of Classical Particles," in
{\em Fundamental Interactions at High Energy}, Coral Gables 1969,
T. Gudehus et al. eds., Gordon and Breach 1969, p.344.
\newline 9. H. van Dam and E.P. Wigner, {\em Phys. Rev.} {\bf 138}, B1576 (1965); {\bf 142}, 838
(1966).
\newline 10. J.A. Wheeler and R.P. Feynman, {\em Rev. Mod Phys.} {\bf 21}, 425 (1949).
\newline 11. D.G. Currie, T.F.
Jordan and E.C.G. Sudarshan, {\em Rev. Mod. Phys.} {\bf 35}, 350
(1963), for the two particle case; H. Leutwyler, {\em Nuovo Cim.}
{\bf XXXVII}, 556 (1965) for the $N$ particle case; D.G. Currie
and T.F. Jordan, "Interactions in relativistic classical particle
mechanics," in (Boulder) Lectures in Theoretical Physics Vol.X-A,
{\em Quantum Theory and Statistical Mechanics}, A.O. Barut and
W.E. Brittin eds., Gordon and Breach 1968, p.91, for a general
review; V.V. Molotkov and I.T. Todorov, {\em Comm. Math. Phys.}
{\bf 79}, 111 (1981), for a proof in the constraint Hamiltonian
formulation.
\newline 12. P.A.M. Dirac, {\em Rev. Mod. Phys.} {\bf 21}, 392 (1949);
E.C.G. Sudarshan, "Structure of Dynamical theories," in 1961
Brandeis Summer Institute {\em Lectures in Theoretical Physics},
Vol.2, Benjamin 1962, esp. Sect.5, p.143; D.G. Currie, T.F. Jordan
and E.C.G. Sudarshan, {\em Rev. Mod. Phys.} {\bf 35}, 350 (1963).
\newline 13.  H. Ekstein, {\em Comm. Math. Phys.} {\bf 1}, 6 (1965).
\newline 14.  W. Heisenberg, {\em The Physical Principles
of the Quantum Theory}, Dover 1930, esp. Chap.\,III.
\newline 15.   L.L. Landau, "On the
quantum theory of fields," in {\em Niels Bohr and the development
of Physics}, W. Pauli et al. eds. McGraw Hill, 1955.
\newline 16. R.F. Streater and A.S. Wightman, {\em
P C T, Spin and Statistics and All That}, Benjamin 1980; R. Jost,
{\em The General Theory of Quantized Fields}, Am. Math. Soc. 1965.
\newline 17. H.
Lehmann, K. Symanzik and W. Zimmermann, {\em Nuovo Cim.} {\bf 1},
1425 (1955); for excellent accounts of the LSZ theory see R.
Hagedorn, {\em Introduction to Field Theory and Dispersion
Relations}, Pergamon Press 1963 and R. Schweber, {\em Introduction
to Relativistic Quantum Field Theory}, Harper and Row 1961, Sect.
18 b.
\newline 18. K. Hepp,
"On the connection between the Wightman and the LSZ quantum field
theory," in Brandeis Summer Inst. Theor. Phys. 1965, Vol.I, {\em
Axiomatic Field theory}, M. Chretien and S. Deser eds., Gordon and
Breach 1966.
\newline 19. R. Haag, {\em Local Quantum Physics. Fields, Particles,
Algebras}, Springer 1996.
\newline 20. R. Jost, {\em The General theory of Quantized Fields},
Am. Math. Soc. 1965, Chap. VI; A.S. Wightman, "Recent achievements
of axiomatic field theory," in {\em Theoretical Physics}, Trieste
1962, IAEA 1963.
\newline 21.  R.J. Eden, P.V. Landshoff
and D.I. Olive, {\em The analytic $S$-matrix}, Cambridge Univ.
Press 1966.
\newline 22.  H. Araki, Ann. Phys. {\bf 11}, 260 (1960).
\newline 23. A. Martin, {\em Scattering Theory: Unitarity,
Analyticity and Crossing }, Lecture Notes in Physics 3, Springer
1969; A. Martin, "The rigorous analyticity-unitarity program: a
historical account," CERN-TH.6894/93, dedicated to the memory of
Misha Polivanov.
\newline 24. K. Gawedzski and A. Kupiainen, {\em Nucl. Phys.} {\bf B257},
474 (1985). Non positive $\ph^4$ theories have been investigated
by Albeverio and collaborators, see {\em Comm. Math. Phys.} {\bf
184}, 509 (1997).
\newline 25. F. Strocchi, "Gauss's law in local quantum
field theory," in {\em Field Theory, Quantization and Statistical
Mechanics}, E. Tirapegui ed., In memory of B. Jouvet, D. Reidel
Publ. 1981, p. 227 and F. Strocchi, {\em Elements of Quantum
Mechanics of Infinite Systems}, World Scientific 1985, Part C, Ch.
II.
\newline 26. F. Strocchi and A.S. Wightman,
{\em Jour. Math. Phys.} {\bf 15}, 2198 (1974); A.S. Wightman, {\em
Nuovo Cim.} {\bf 110B}, 751, (1995).
\newline 27. R. Ferrari, L.E. Picasso and F. Strocchi, {\em Comm.
Math. Phys.} {\bf 35}, 25 (1974).
\newline 28. F.
Strocchi, {\em Phys. Rev.} {\bf D17}, 2010 (1978); F. Strocchi,
{\em Selected Topics on the General Properties of Quantum Field
Theory}, World Scientific 1993, esp. Ch. VI.
\newline 29. S. Doplicher,
R. Haag and J. Roberts, {\em Comm. Math. Phys.} {\bf 13}, 1,
(1969); {\em ibid} {\bf 15}, 173 (1969); R. Haag, {\em Local
Quantum Physics}, Springer 1996, Ch. IV, Sect.2.
\newline 30.  G. Morchio
and F. Strocchi, lectures at the Erice School on {\em Fundamental
Problems of Gauge Field Theory}, G. Velo and A.S. Wightman eds.,
Plenum Press 1986; G. Morchio and F. Strocchi,  {\em Jour. Math.
Phys.} {\bf 44}, 5569 (2003).
\newline 31. A. Bassetto, G. Nardelli and R. Soldati, {\em
Yang-Mills Theories in Algebraic Non-covariant Gauges}, World
Scientific 1991, esp. Sect.3.3.
\newline 32. J.
Loeffelholz, G. Morchio and F. Strocchi,  {\em Journ. Math. Phys.}
{\bf 44}, 5095 (2003).
\newline 33. F. Acerbi, G. Morchio and F, Strocchi, {\em Jour. Math. Phys.} {\bf 34},
899 (1993).
\newline 34. S. Cavallaro, G. Morchio and F. Strocchi, {\em Lett. Math. Phys.} {\bf
47}, 307 (1999).
\newline 35. F. Acerbi, G. Morchio and F. Strocchi, {\em  Lett.
Math. Phys.} {\bf 27}, 1 (1992); {\em Reports Math. Phys.} {\bf
33}, 7 (1993); J. Loeffelholz, G. Morchio and F. Strocchi, {\em
Lett. Math. Phys.} {\bf 35}, 251 (1995).
\newline 36.  F. Acerbi, G. Morchio and F. Strocchi, {\em Lett. Math.
Phys.} {\bf 26}, 13 (1992); ibid. {\bf 27} 1 (1993); J.
Loeffelholz, G. Morchio and F. Strocchi, {\em Ann. Phys.} {\bf
250}, 367 (1996); for the Chern-Simons models see F. Nill, {\em
Int. Jour. Mod. Phys.} B {\bf 6} 2159 (1992).
\newline 37.  G. Morchio and F.
Strocchi, Erice lectures 1985, loc. cit.; {\em Journ. Math. Phys.}
{\bf 28} 622) (1987); ibid. {\bf 28} 1912 (1987); Comm. Math.
Phys. {\bf 111}, 593 (1987); "Removal of the infrared cutoff,
seizing of the vacuum and symmetry breaking in many body and in
gauge theories," Invited talk at the {\em IX International
Congress on Mathematical Physics}, Swansea 1988, B. Simon et al.
eds., Adam Hilger 1989, p.\,490.
\newline 38.  G. Morchio and F. Strocchi,  {\em
Jour. Math. Phys.} {\bf 44}, 5569 (2003).
\end{document}